\documentclass[aps,pra,twocolumn,tightenlines, footinbib, longbibliography,showkeys,showpacs]{revtex4-2} 
\usepackage{graphicx}
\usepackage{dcolumn}
\usepackage{bm}
\usepackage[colorlinks,linkcolor=blue,anchorcolor=blue,citecolor=blue,urlcolor=blue]{hyperref}
\usepackage{color}
\usepackage{amsmath}
\usepackage{amsthm}
\usepackage{amssymb}
\usepackage{mathrsfs}
\begin{document}
\title{Anti-$\mathcal{PT}$-symmetric Kerr gyroscope}
\author{Huilai Zhang}
\affiliation{Key Laboratory of Low-Dimensional Quantum Structures and Quantum Control of Ministry of Education,\\
Department of Physics and Synergetic Innovation Center for Quantum Effects and Applications,\\
Hunan Normal University, Changsha 410081, China}

\author{Meiyu Peng}
\affiliation{Key Laboratory of Low-Dimensional Quantum Structures and Quantum Control of Ministry of Education,\\
Department of Physics and Synergetic Innovation Center for Quantum Effects and Applications,\\
Hunan Normal University, Changsha 410081, China}

\author{Xun-Wei Xu}
\affiliation{Key Laboratory of Low-Dimensional Quantum Structures and Quantum Control of Ministry of Education,\\
Department of Physics and Synergetic Innovation Center for Quantum Effects and Applications,\\
Hunan Normal University, Changsha 410081, China}

\author{Hui Jing}
\email[Corresponding author. ]{jinghui73@foxmail.com}
\affiliation{Key Laboratory of Low-Dimensional Quantum Structures and Quantum Control of Ministry of Education,\\
Department of Physics and Synergetic Innovation Center for Quantum Effects and Applications,\\
Hunan Normal University, Changsha 410081, China}

\date{\today}

\begin{abstract}
Non-Hermitian systems can exhibit unconventional spectral singularities called exceptional points (EPs). Various EP sensors have been fabricated in recent years, showing strong spectral responses to external signals. Here we propose how to achieve a nonlinear anti-parity-time ($\mathcal{APT}$) gyroscope by spinning an optical resonator. We show that, in the absence of any nonlinearity, the sensitivity or optical mode splitting of the linear device can be magnified up to 3 orders than that of the conventional device without EPs. Remarkably, the $\mathcal{APT}$ symmetry can be broken when including the Kerr nonlinearity of the materials and, as the result, the detection threshold can be significantly lowered, i.e., much weaker rotations which are well beyond the ability of a linear gyroscope can now be detected with the nonlinear device. Our work shows the powerful ability of $\mathcal{APT}$ gyroscopes in practice to achieve ultrasensitive rotation measurement.
\end{abstract}


\maketitle
\section{Introduction}
\begin{figure*}[htbp]
	\includegraphics[width=0.85 \textwidth]{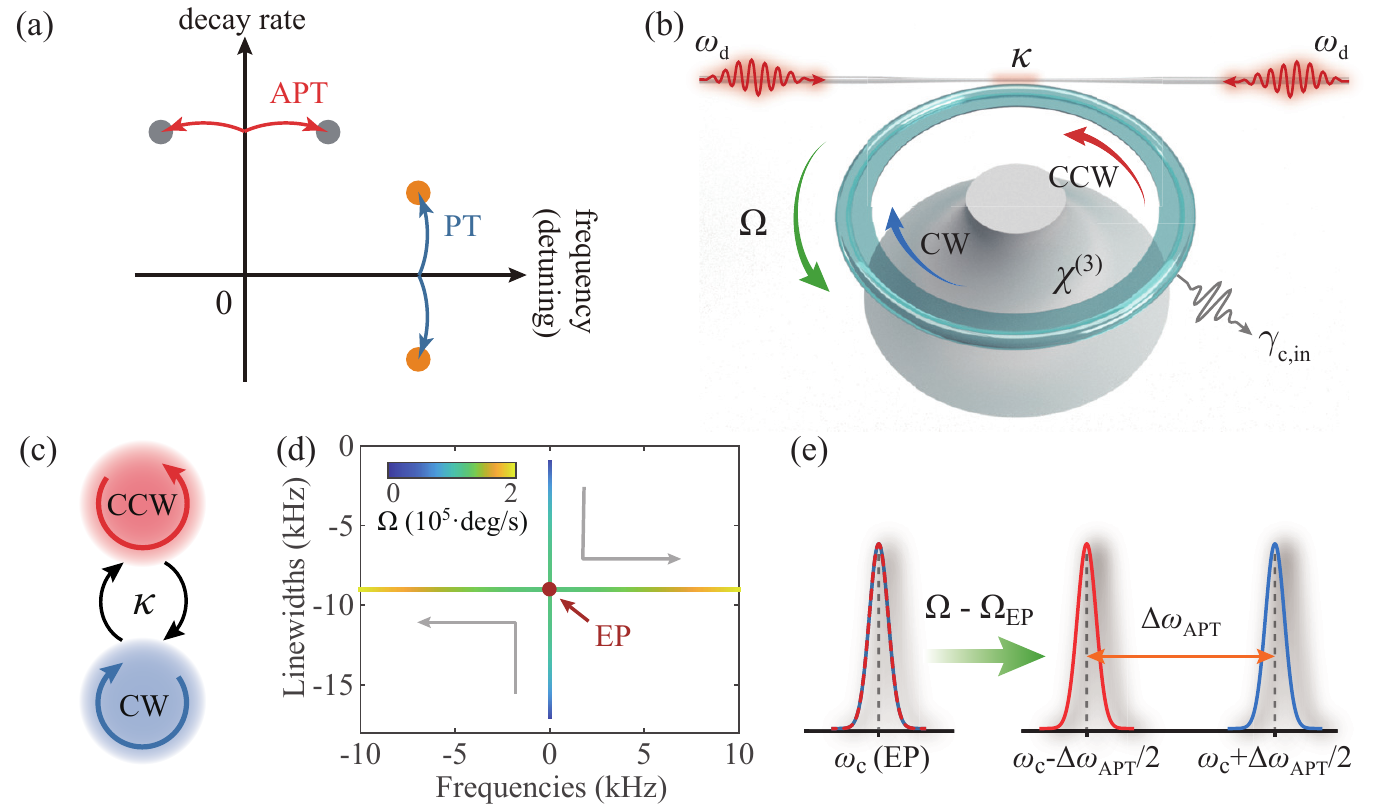}
	\caption{\label{Fig1} Anti-parity-time ($\mathcal{APT}$)-symmetric gyroscope. (a) Comparison between $\mathcal{PT}$ and $\mathcal{APT}$ symmetries. (b) Schematic of a nonlinear $\mathcal{APT}$ gyroscope based on an optical spinning resonator. (c) Schematic of dissipatively coupled counter-propagating modes. (d) Linear eigenfrequencies versus rotation speed $\Omega$ in the complex plane. (e) Mechanism of $\mathcal{APT}$ gyroscope. }
\end{figure*}

Exceptional points (EPs)~\cite{miri2019Exceptional} are non-Hermitian spectral degeneracies, where both the eigenvalues and the corresponding eigenvectors of a system simultaneously coalesce. Nontrivial phenomena related to EPs have been of great interest, particularly in connection with parity-time ($\mathcal{PT}$)~\cite{bender1998Real,zyablovsky2014PTsymmetry,konotop2016Nonlinear,longhi2017Paritytime,feng2017NonHermitian,el-ganainy2018NonHermitian,ozdemir2019Parity} and anti-$\mathcal{PT}$ ($\mathcal{APT}$) symmetries~\cite{ge2013antisymmetric,peng2016Antiparitytime}.
$\mathcal{PT}$-symmetric systems have balanced loss and gain, and can exhibit EP with entirely real eigenvalues. Novel applications of $\mathcal{PT}$ systems such as $\mathcal{PT}$-symmetric optomechanics~\cite{jing2014PTSymmetric,zhang2018Phonon}, topological photonics~\cite{zhang2018Dynamically,yoon2018Timeasymmetric}, $\mathcal{PT}$ metamaterials~\cite{feng2013Experimental,castaldi2013PT} to name only a few, have been revealed.
On the contrary, $\mathcal{APT}$ symmetry does not need gain but can still exhibit EP with entirely imaginary eigenvalues in the symmetric phase (see Fig.~\ref{Fig1}(a))~\cite{ge2013antisymmetric}, which has been extensively explored both in experiments~\cite{peng2016Antiparitytime,jiang2019AntiParityTime,fan2020AntiparityTime,zhang2019Dynamically,park2021Optical,zhao2020Observation,yang2020Unconventional,li2019Antiparitytime,xu2021Configurable,choi2018Observation,li2019Experimental,ding2021Information,bergman2021Observation,wen2020Observation,cao2020ReservoirMediated,zhang2020Synthetic,stegmaier2021Topological,wang2020AllOptical} and in theories~\cite{yang2017AntiPT,dai2020Asymmetric,lu2019Berry,zhang2020Breaking,xu2021Couplinginduced,zheng2019Duality,wang2020Effective,qi2021Encircling,nair2021Enhanced,li2021GainFree,cao2020HighOrder,carlo2019Highsensitivity,abbas2020Investigation,peng2020Level,shui2019Lopsided,roy2021Nondissipative,wu2014nonhermitian,konotop2018OddTime,wang2016Optical,ryu2019Oscillation,wu2015Paritytimeantisymmetric,antonosyan2015Paritytime,chen2017PseudoHermitian,longhi2018PT,zhao2021Realpotentialdriven,jin2018Scattering,duan2021Singlecavity,chen2020Spontaneous,ke2019Topological,wu2021Topology,jahromi2021Witnessing}. This striking difference provides us a variety of novel opportunities to explore $\mathcal{APT}$ effects in atomic systems~\cite{peng2016Antiparitytime,jiang2019AntiParityTime}, photonic or magnonic devices~\cite{fan2020AntiparityTime,zhang2019Dynamically,park2021Optical,zhao2020Observation,yang2020Unconventional}, and thermal structures~\cite{li2019Antiparitytime,xu2021Configurable}. These systems are known to exhibit nontrivial phenomena such as chiral mode switching~\cite{zhang2019Dynamically}, energy-difference conserving dynamics~\cite{choi2018Observation,park2021Optical}, and constant refraction~\cite{yang2017AntiPT}.
In particular, EP systems, including those with $\mathcal{PT}$ and $\mathcal{APT}$ symmetries, can open up a new route to make ultrasensitive sensors since they can display a strong response to perturbations near EPs~\cite{wiersig2014Enhancing,wiersig2020Review}.

As highly sensitive rotation sensors, gyroscopes~\cite{li2017Microresonator,li2016Rotation,li2018Improving} are widely used in e.g. unmanned and autonomous vehicles and global navigation systems. The concept of EP sensing applied to ultrasensitive rotation measurement was first proposed in a $\mathcal{PT}$ system~\cite{ren2017Ultrasensitive}. After that, several EP-based gyroscopes have been proposed using $\mathcal{PT}$ or $\mathcal{APT}$ symmetry~\cite{smith2019Paritytimesymmetrybreaking,grant2020Enhanced,Mao2020,carlo2019Highsensitivity}, loss regulation~\cite{sunada2017Large,Grant2021}, and dissipative coupling~\cite{lai2020Earth,wang2020Petermannfactor,horstman2020Exceptional}. Compared with $\mathcal{PT}$ gyroscope, $\mathcal{APT}$ gyroscope does not need gain, thus can be kept at EP more accurately. Moreover, $\mathcal{APT}$ gyroscope can exhibit a completely real frequency splitting which can be directly measured at the output power spectrum~\cite{carlo2019Highsensitivity}, while $\mathcal{PT}$ gyroscope usually exhibits complex frequency splitting.
In addition, in two recent experiments, EP-based quasi-$\mathcal{APT}$~\cite{lai2019Observation} and passive-$\mathcal{PT}$~\cite{hokmabadi2019NonHermitian} gyroscopes were demonstrated with a microresonator and a macroscopic triangular cavity, respectively.
However, these studies mainly focused on the linear case. The effect of nonlinearity on EP gyroscope remains largely unclear, despite a few discussions along this line~\cite{wang2020Petermannfactor,lai2019Observation,hokmabadi2019NonHermitian}. Given that the superior performance of linear EP gyroscope has been confirmed in the experiments~\cite{lai2019Observation,hokmabadi2019NonHermitian}, it is thus necessary to go a step further to explore the role of nonlinearity in $\mathcal{APT}$ gyroscope.

In a recent experiment~\cite{maayani2018Flying}, a spinning resonator was used for realizing nonreciprocal light transmission with $99.6\%$ rectification. Such spinning resonators~\cite{huang2018Nonreciprocal,li2019Nonreciprocal,jing2018Nanoparticle,jiao2020Nonreciprocal,lu2017Optomechanically,li2021Nonreciprocal,xu2021Nonreciprocal,jiang2018Nonreciprocal,li2020Nonreciprocal} have also been applied to single-photon control~\cite{huang2018Nonreciprocal,li2019Nonreciprocal}, single nanoparticles sensing~\cite{jing2018Nanoparticle}, and backscattering immune optomechanical entanglement generation~\cite{jiao2020Nonreciprocal} to name a few. We also find that~\cite{zhang2020Breaking} the opposite Sagnac-Fizeau shifts in counter-propagating modes of the resonator can be utilized for creating an $\mathcal{APT}$-symmetric system.
Moreover, this system is linear and can be readily extended to a system with nonlinear materials.

To explore the role of nonlinearity in $\mathcal{APT}$ gyroscope, in this work, we propose a nonlinear $\mathcal{APT}$-symmetric optical gyroscope based on a spinning resonator with Kerr nonlinearity of the materials.
To have a better comparison, we examine the $\mathcal{APT}$ gyroscope without nonlinearity first, and show that the sensitivity of frequency splitting can be magnified $4.9\times10^{3}$ times that of a conventional device without EPs.
Then, in presence of Kerr nonlinearity~\cite{cao2017Experimental,delbino2017Symmetry,cao2020Reconfigurable} of materials,
we find that the sensitivity can be further enhanced in $\mathcal{APT}$-symmetry-broken ($\mathcal{APT}$B) regime, which is $\mathcal{APT}$ symmetric ($\mathcal{APT}$S) in the linear device. The nonlinearity-induced phase transition is reminiscent of the similar effects in $\mathcal{PT}$ systems~\cite{konotop2016Nonlinear,lumer2013Nonlinearly}. Remarkably, the detection threshold of the $\mathcal{APT}$ gyroscope can be significantly lowered, i.e., much weaker rotations which are well beyond the ability of a linear device can now be detected with the nonlinear one.
The parameters used for numerical simulations are experimentally feasible~\cite{maayani2018Flying,lai2019Observation,huet2016Millisecond,shen2016Compensation,brasch2016Photonic}. Our work shows the powerful ability of $\mathcal{APT}$ gyroscopes in practice to achieve ultrasensitive rotation measurement.

\section{$\mathcal{APT}$-symmetric gyroscope}
\subsection{Model and Hamiltonian}
As schematically shown in Fig.~\ref{Fig1}(b), we propose a nonlinear $\mathcal{APT}$ gyroscope using an optical whispering-gallery-mode resonator. The resonator is pumped bidirectionally at frequency $\omega_{\mathrm{d}}$ with amplitude $\varepsilon_{\mathrm{d}} = \sqrt{\gamma_{\mathrm{ex}}P/\hbar\omega_{\mathrm{d}}}$ ($\gamma_{\mathrm{ex}}$ is the coupling-induced loss and $P$ is the input power) to excite two counter-propagating modes, i.e., clockwise (CW) and counter-clockwise (CCW) modes. The resonant frequency and intrinsic optical loss of them are  $\omega_{\mathrm{c}}$ and $\gamma_{\mathrm{c,in}}$, respectively. A tapered fiber is evanescently coupled to the resonator, resulting in dissipative coupling between CW and CCW modes~\cite{lai2019Observation}, as illustrated in Fig.~\ref{Fig1}(c).
In presence of self-Kerr and cross-Kerr nonlinearities~\cite{cao2017Experimental,delbino2017Symmetry,cao2020Reconfigurable} of the materials and in the rotating frame at the pump frequency,  the nonlinear Hamiltonian can be expressed as ($\hbar=1$)~\cite{lai2019Observation,zhang2020Breaking}
\begin{small}
\begin{align}\label{eq:NonlinearHamiltonian}
	H_{\mathrm{NL}} & =\left(\Delta_{\mathrm{c}} - i\gamma_{\mathrm{c}}\right)\left(a_{\mathcal{\mathrm{\circlearrowright}}}^{\dagger}a_{\mathrm{\circlearrowright}} +a_{\mathcal{\mathrm{\circlearrowleft}}}^{\dagger}a_{\mathrm{\circlearrowleft}}\right)\nonumber \\
	& \quad+i\kappa\left(a_{\mathcal{\mathrm{\circlearrowright}}}^{\dagger}a_{\mathrm{\circlearrowleft}}+a_{\mathcal{\mathrm{\circlearrowleft}}}^{\dagger}a_{\mathrm{\circlearrowright}}\right) +g\left(a_{\mathcal{\mathrm{\circlearrowright}}}^{\dagger2}a^{2}_{\mathrm{\circlearrowright}}+a_{\mathcal{\mathrm{\circlearrowleft}}}^{\dagger2}a^{2}_{\mathrm{\circlearrowleft}}\right)\nonumber \\
	& \quad+4ga_{\mathcal{\mathrm{\circlearrowright}}}^{\dagger}a_{\mathrm{\circlearrowright}}a_{\mathcal{\mathrm{\circlearrowleft}}}^{\dagger}a_{\mathrm{\circlearrowleft}} +i\varepsilon_{\mathrm{d}}\left(a_{\mathrm{\circlearrowright}}^{\dagger}-a_{\mathrm{\circlearrowright}}+a_{\mathrm{\circlearrowleft}}^{\dagger}-a_{\mathrm{\circlearrowleft}}\right),
\end{align}
\end{small}
where $$\Delta_{\mathrm{c}}=\omega_{\mathrm{c}}-\omega_{\mathrm{d}},~~ \gamma_{\mathrm{c}} = (\gamma_\mathrm{ex} + \gamma_{\mathrm{c,in}})/2,$$ and $\kappa$ is the dissipative coupling strength. $a_{\mathrm{\circlearrowright}}$ ($a^{\dagger}_{\mathrm{\circlearrowright}}$) and $a_{\mathrm{\circlearrowleft}}$ ($a^{\dagger}_{\mathrm{\circlearrowleft}}$) are the annihilation (creation) operators of CW and CCW modes.
$g = \hbar\omega_{\mathrm{c}}^2cn_{\mathrm{NL}}/n^2V_{\mathrm{eff}}$ is the coefficient of Kerr nonlinearity~\cite{shen2016Compensation,brasch2016Photonic} where $n$ and $n_{\mathrm{NL}}$ are the linear and nonlinear refractive index of materials,  $V_{\mathrm{eff}}$ is the effective mode volume of the resonator, and $c$ is the speed of light in the vacuum.
\begin{figure}[tbp]
	\includegraphics[width=0.9 \columnwidth]{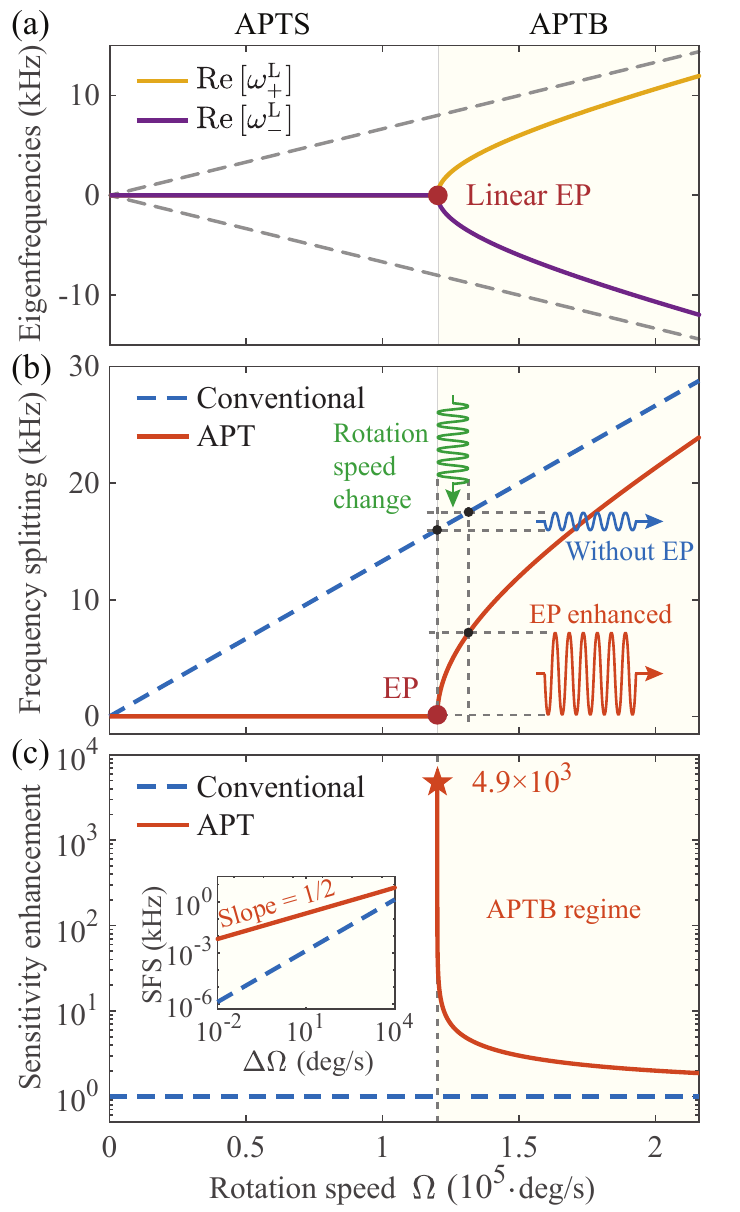}
	\caption{\label{Fig2} Linear $\mathcal{APT}$ gyroscope. (a) Real parts of eigenfrequencies of conventional (dashed curves) and $\mathcal{APT}$ (solid curves) gyroscopes versus the rotation speed $\Omega$. (b) Frequency splitting versus the rotation speed $\Omega$ in the conventional (blue dashed curve) and $\mathcal{APT}$ (orange solid curve) gyroscopes. (c) Sensitivity enhancement versus the rotation speed $\Omega$ (orange solid curve). The blue dashed curve is corresponding to the conventional gyroscope. The inset shows the shifted frequency splitting (SFS) versus the change of rotation speed $\Delta\Omega = \Omega - \Omega_{\mathrm{EP}}$.}
\end{figure}

When this resonator of radius $R$ rotates CCW at a speed $\Omega$, the CW and
CCW modes will experience opposite Sagnac-Fizeau shifts $\Delta_{\mathrm{c}}\rightarrow\Delta_{\mathrm{c}}\pm\Delta_{\mathrm{sag}}$ where~\cite{malykin2000Sagnac}
\begin{equation}\label{eq:Sagnac}
	\Delta_{\mathrm{sag}}=\frac{n R\Omega\omega_{\mathrm{c}}}{c}\left(1-\frac{1}{n^{2}}-\frac{\lambda}{n}\frac{\mathrm{d}n}{\mathrm{d}\lambda}\right),
\end{equation}
where the dispersion term $\mathrm{d}n/\mathrm{d}\lambda$ characterizing the relativistic origin of the Sagnac effect is relatively small ($\sim1\%$) in typical materials of microresonator. Then the Hamiltonian of the spinning gyroscope can be obtained as
\begin{small}
\begin{align}\label{eq:FullNonlinearHamiltonian}
	H_{\mathrm{NL}} & =\left(\Delta_{\mathrm{c}} + \Delta_{\mathrm{sag}}- i\gamma_{\mathrm{c}}\right)a_{\mathcal{\mathrm{\circlearrowright}}}^{\dagger}a_{\mathrm{\circlearrowright}}+\left(\Delta_{\mathrm{c}} - \Delta_{\mathrm{sag}}-i\gamma_{\mathrm{c}}\right)a_{\mathcal{\mathrm{\circlearrowleft}}}^{\dagger}a_{\mathrm{\circlearrowleft}}\nonumber \\
	& \quad+i\kappa\left(a_{\mathcal{\mathrm{\circlearrowright}}}^{\dagger}a_{\mathrm{\circlearrowleft}}+a_{\mathcal{\mathrm{\circlearrowleft}}}^{\dagger}a_{\mathrm{\circlearrowright}}\right) +g\left(a_{\mathcal{\mathrm{\circlearrowright}}}^{\dagger2}a^{2}_{\mathrm{\circlearrowright}}+a_{\mathcal{\mathrm{\circlearrowleft}}}^{\dagger2}a^{2}_{\mathrm{\circlearrowleft}}\right)\nonumber \\
	& \quad+4ga_{\mathcal{\mathrm{\circlearrowright}}}^{\dagger}a_{\mathrm{\circlearrowright}}a_{\mathcal{\mathrm{\circlearrowleft}}}^{\dagger}a_{\mathrm{\circlearrowleft}} +i\varepsilon_{\mathrm{d}}\left(a_{\mathrm{\circlearrowright}}^{\dagger}-a_{\mathrm{\circlearrowright}}+a_{\mathrm{\circlearrowleft}}^{\dagger}-a_{\mathrm{\circlearrowleft}}\right).
\end{align}
\end{small}
We use the following parameters in numerical simulations: $\lambda = 1550\,\mathrm{nm}$, $Q \approx 2\times10^{10}$, $\gamma_{\mathrm{ex}} = \gamma_{\mathrm{c,in}}$, $\kappa = 8\,\mathrm{kHz}$, $n = 1.44$, $R = 50\,\mathrm{\mu m}$, as in relevant experiments~\cite{maayani2018Flying,lai2019Observation,huet2016Millisecond}. The coefficient of Kerr nonlinearity $g$ is chosen to be $0.17\,\mathrm{Hz}$ according to the experimentally feasible parameters~\cite{shen2016Compensation,brasch2016Photonic}: $n_{\mathrm{NL}} = 3.2\times10^{-16}\,\mathrm{cm^{2}/W}$ and $V_{\mathrm{eff}} = 10^{2}\,\mathrm{\mu m^3}$.

\subsection{Linear $\mathcal{APT}$ gyroscope}
In this part, we will first study the spinning $\mathcal{APT}$ gyroscope in absence of nonlinearity.
According to Eq.~(\ref{eq:FullNonlinearHamiltonian}), the linear Hamiltonian
without the driving terms reads
\begin{equation}
	H_{\mathrm{L}} =\left(\begin{array}{cc}
		a_{\mathcal{\mathrm{\circlearrowright}}}^{\dagger} & a_{\mathrm{\circlearrowleft}}^{\dagger}\end{array}\right)M_{\mathrm{L}}\left(\begin{array}{c}
		a_{\mathrm{\mathrm{\circlearrowright}}}\\
		a_{\mathrm{\mathrm{\circlearrowleft}}}
	\end{array}\right),
\end{equation}
with a $2\times2$ coefficient matrix
\begin{equation}\label{eq:EffectiveLinearHamiltonian}
	M_{\mathrm{L}} = \left(\begin{array}{cc}
		\Delta_{\mathrm{c}}+\Delta_{\mathrm{sag}}-i\gamma_{\mathrm{c}} & i\kappa\\
		i\kappa & \Delta_{\mathrm{c}}-\Delta_{\mathrm{sag}}-i\gamma_{\mathrm{c}}
	\end{array}\right).
\end{equation}
When $\Delta_{\mathrm{c}} = 0$, the Hamiltonian is antisymmetric under combined $\mathcal{PT}$ operation. According to Eq.~(\ref{eq:EffectiveLinearHamiltonian}) and $\left|M_{\mathrm{L}}-\omega\mathbb{I}_{2}\right|=0$ ($\mathbb{I}_{2}$ is a $2\times2$ identity matrix), the eigenfrequencies are derived as
\begin{align}\label{eq:LinearEigenfrequencies}
	\omega^{\mathrm{L}}_{\pm} & =-i\gamma_{\mathrm{c}}\pm\sqrt{\Delta_{\mathrm{sag}}^{2}-\kappa^{2}}.
\end{align}
The evolution of eigenfrequencies in the complex plane is plotted as a function of the rotation speed $\Omega$ in Fig.~\ref{Fig1}(d). As depicted in Fig.~\ref{Fig1}(d), the eigenfrequency splitting is purely imaginary in $\mathcal{APT}$S regime before EP. When $\Delta_{\mathrm{sag}} = \kappa$, i.e., $\Omega = \Omega^{\mathrm{L}}_{\mathrm{EP}} \approx 1.2\times10^5\,\mathrm{deg/s}$, this system is at EP where the real and imaginary parts of the eigenfrequencies simultaneously coalesce. By further increasing the rotation speed, a real splitting of eigenfrequencies can be observed in $\mathcal{APT}$B regime. Thus, this spinning $\mathcal{APT}$ resonator can function as a gyroscope for rotation measurement around EP by breaking $\mathcal{APT}$ symmetry, as illustrated in Fig.~\ref{Fig1}(e).

In Fig.~\ref{Fig2}(a), we plot the real parts of eigenfrequencies $\mathrm{Re}\,[\omega^{\mathrm{L}}_{\pm}]$ versus rotation speed $\Omega$. A conventional gyroscope is shown for comparison, in which the real parts of eigenfrequencies is proportional to rotation speed, see the dashed curves in Fig.~\ref{Fig2}(a).
For $\mathcal{APT}$ gyroscope, the frequencies are locked in $\mathcal{APT}$S regime before EP, creating a measuring dead band for rotations~\cite{lai2019Observation,horstman2020Exceptional}. While in the $\mathcal{APT}$B regime (shaded regimes in Fig.~\ref{Fig2}), the rotation speed over $\Omega^{\mathrm{L}}_{\mathrm{EP}}$ leads to a splitting between the unlocked eigenfrequencies, which is
\begin{align}
	\Delta\omega_{\mathrm{L}} & =\mathrm{Re}\left[\omega^{\mathrm{L}}_{+}-\omega^{\mathrm{L}}_{-}\right]=2\mathrm{Re}\left[\sqrt{\Delta_{\mathrm{sag}}^{2}-\kappa^{2}}\right].
\end{align}
Figure~\ref{Fig2}(b) provides a comparison between $\mathcal{APT}$ and conventional gyroscopes, from which the frequency splitting of $\mathcal{APT}$ gyroscope can be observed in the $\mathcal{APT}$B regime. For the conventional gyroscope,  a frequency splitting emerges once a rotation is applied to the system.

To characterize the performance of $\mathcal{APT}$ gyroscope, we define a sensitivity enhancement factor as $\eta_{\mathrm{L}} = \Lambda_{\mathrm{L}}/\Lambda_{\mathrm{CONV}}$, with $\Lambda_{\mathrm{L}} = \Delta\omega_{\mathrm{L}}/\Delta\Omega$ and $\Lambda_{\mathrm{CONV}} = \Delta\omega_{\mathrm{CONV}}/\Delta\Omega$. The frequency splittings $\Delta\omega_{\mathrm{L}}$ and $\Delta\omega_{\mathrm{CONV}}$ are in respect to change of rotation speed $\Delta\Omega = \Omega - \Omega^{\mathrm{L}}_{\mathrm{EP}}$ (illustrated in Fig.~\ref{Fig2}(b)), given that the frequency splitting of $\mathcal{APT}$ gyroscope only emerges in $\mathcal{APT}$B regime. The inset in Fig.~\ref{Fig2}(c) shows $\Delta\omega_{\mathrm{L}}$ and $\Delta\omega_{\mathrm{CONV}}$ versus $\Delta\Omega$ in log-log scale. The slope $1/2$ of the curve corresponding to $\mathcal{APT}$ gyroscope depicts that the frequency splitting has a square-root dependence on $\Delta\Omega$, indicating much more enhancement of frequency splitting for smaller $\Delta\Omega$. The main plot in Fig.~\ref{Fig2}(c) shows the sensitivity enhancement versus $\Omega$. When experiencing a detectable change of rotation speed of $0.01\,\mathrm{deg/s}$~\cite{lai2020Earth} from EP in $\mathcal{APT}$B regime, the sensitivity of frequency splitting in $\mathcal{APT}$ gyroscope can be magnified $4.9\times10^3$ times that of the conventional gyroscope without EPs (see the orange solid curve in the main plot in Fig.~\ref{Fig2}(c)). Although the overall frequency splitting of conventional gyroscope is larger than that of $\mathcal{APT}$ gyroscope, the conventional gyroscope exhibits no enhancement in measuring rotation speed, see the blue dashed curve in the main plot in Fig.~\ref{Fig2}(c).
\begin{figure}[tbp]
	\includegraphics[width=0.9 \columnwidth]{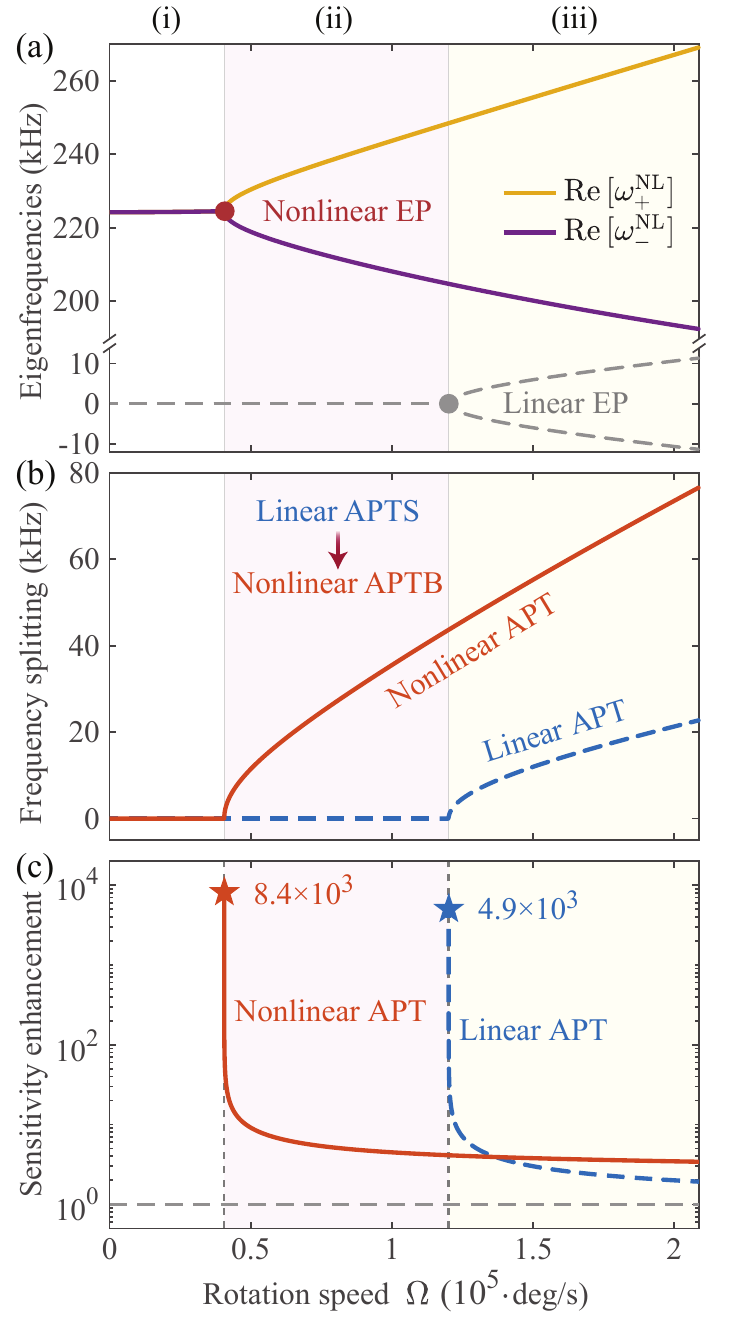}
	\caption{\label{Fig3} Nonlinear $\mathcal{APT}$ gyroscope. (a) Real parts of eigenfrequencies of the nonlinear (solid curves) and linear (dashed curves) $\mathcal{APT}$ gyroscopes versus the rotation speed $\Omega$. (b) Frequency splittings of the nonlinear (orange solid curve) and linear (blue dashed curve) $\mathcal{APT}$ gyroscopes versus $\Omega$. (c) The nonlinear (orange solid curve) and linear (blue dashed curve) sensitivity enhancements versus $\Omega$. $g$ is chosen to be $0.17\,\mathrm{Hz}$ according to the experimentally feasible parameters~\cite{shen2016Compensation,brasch2016Photonic}: $n_{\mathrm{NL}} = 3.2\times10^{-16}\,\mathrm{cm^{2}/W}$ and $V_{\mathrm{eff}} = 10^{2}\,\mathrm{\mu m^3}$. The pump power is $P = 25\,\mathrm{nW}$.}
\end{figure}

\subsection{Nonlinear $\mathcal{APT}$ gyroscope}
We have confirmed that the frequency splitting can be enhanced in the linear $\mathcal{APT}$ gyroscope above.
Now we explore the role of nonlinearity in $\mathcal{APT}$ gyroscope. According to Eq.~(\ref{eq:FullNonlinearHamiltonian}), the equations of motion can be derived as
\begin{align}\label{eq:NonlinearEOMCW}
	\dot{a}_{\mathrm{\circlearrowright}} & =-i\left(\Delta_{\mathrm{c}}+ \Delta_{\mathrm{sag}}-i\gamma_{\mathrm{c}}\right)a_{\mathrm{\circlearrowright}}-i2ga_{\mathcal{\mathrm{\circlearrowright}}}^{\dagger}a_{\mathrm{\circlearrowright}}a_{\mathrm{\circlearrowright}}\nonumber \\
	& \quad-i4ga_{\mathrm{\circlearrowright}}a_{\mathcal{\mathrm{\circlearrowleft}}}^{\dagger}a_{\mathrm{\circlearrowleft}}+\kappa a_{\mathcal{\mathrm{\circlearrowleft}}}+\varepsilon_{\mathrm{d}},\\
	\dot{a}_{\mathrm{\circlearrowleft}} & =-i\left(\Delta_{\mathrm{c}}- \Delta_{\mathrm{sag}}-i\gamma_{\mathrm{c}}\right)a_{\mathrm{\circlearrowleft}}-i2ga_{\mathcal{\mathrm{\circlearrowleft}}}^{\dagger}a_{\mathrm{\circlearrowleft}}a_{\mathrm{\circlearrowleft}}\nonumber \\
	& \quad-i4ga_{\mathrm{\circlearrowleft}}a_{\mathcal{\mathrm{\circlearrowright}}}^{\dagger}a_{\mathrm{\circlearrowright}}+\kappa a_{\mathcal{\mathrm{\circlearrowright}}}+\varepsilon_{\mathrm{d}}. \label{eq:NonlinearEOMCCW}
\end{align}
Taking the expectation amplitudes of operators~\cite{li2021Nonlineardissipationinduced}, i.e., ${\alpha}_{\mathrm{\{\circlearrowright,\circlearrowleft\}}} = \langle {a}_{\mathrm{\{\circlearrowright,\circlearrowleft\}}} \rangle$, Eq.~(\ref{eq:NonlinearEOMCW}) and Eq.~(\ref{eq:NonlinearEOMCCW}) can be rewritten as
\begin{equation}\label{eq:effectiveEOMs}
	\left(\begin{array}{c}
		\dot{\alpha}_{\mathrm{\circlearrowright}}\\
		\dot{\alpha}_{\mathrm{\circlearrowleft}}
	\end{array}\right)=-iM_{\mathrm{NL}}\left(\begin{array}{c}
		\alpha_{\mathrm{\circlearrowright}}\\
		\alpha_{\mathrm{\circlearrowleft}}
	\end{array}\right)+\left(\begin{array}{c}
		\varepsilon_{\mathrm{d}}\\
		\varepsilon_{\mathrm{d}}
	\end{array}\right),
\end{equation}
with the nonlinear coefficient matrix
\begin{small}
\begin{equation}
	M_{\mathrm{NL}}\!=\!\left(\begin{array}{cc}
		\Delta_{\mathrm{c}}\!+\! \Delta_{\mathrm{sag}}\!-\!i\gamma_{\mathrm{c}}\!+\!G_{\mathrm{\circlearrowright}} & i\kappa\\
		i\kappa & \Delta_{\mathrm{c}}\!-\! \Delta_{\mathrm{sag}}\!-\!i\gamma_{\mathrm{c}}\!+\!G_{\mathrm{\circlearrowleft}}
	\end{array}\right)\!,
\end{equation}
\end{small}
where $G_{\mathrm{\circlearrowright, \circlearrowleft}} = 2g\left|\alpha_{\mathrm{\circlearrowright, \circlearrowleft}}\right|^{2}+4g\left|\alpha_{\mathrm{\circlearrowleft, \circlearrowright}}\right|^{2}$.

From $\left|M_{\mathrm{NL}}-\omega\mathbb{I}_{2}\right|=0$, the characteristic equation can be obtained and the eigenfrequencies of the nonlinear system can consequently obtained as
\begin{align}
	\omega_{\pm}^{\mathrm{NL}} & =\Delta_{\mathrm{c}}-i\gamma_{\mathrm{c}}+3g\left(\left|\alpha_{\mathrm{\circlearrowright}}\right|^{2}+\left|\alpha_{\mathrm{\circlearrowleft}}\right|^{2}\right)\nonumber \\
	& \quad\pm\sqrt{\left[g\left(\left|\alpha_{\mathrm{\circlearrowleft}}\right|^{2}-\left|\alpha_{\mathrm{\circlearrowright}}\right|^{2}\right)+\Delta_{\mathrm{sag}}\right]^{2}-\kappa^{2}}.
\end{align}
This equation indicates that nonlinearity can alter the location of EP~\cite{Ramezanpour2021}. By setting $\dot{\alpha}_{\mathrm{\{\circlearrowright,\circlearrowleft\}}} = 0$ and $\Delta_{\mathrm{c}} = 0$,  the values of the formal solutions of eigenfrequencies can be obtained by numerically solving Eq.~(\ref{eq:effectiveEOMs}). Figure~\ref{Fig3}(a) plots the real parts of eigenfrequencies as a function of rotation speed $\Omega$. The presence of nonlinearity results in a lower rotation speed corresponding to EP (illustrated by the red dot) which is $\Omega^{\mathrm{NL}}_{\mathrm{EP}} \approx 4.06 \times 10^{4}\,\mathrm{deg/s}$ for $g = 0.17\,\mathrm{Hz}$, leading to a narrower measuring dead band than the linear gyroscope.
We can see in regime (i), the linear and nonlinear gyroscopes both are $\mathcal{APT}$S with locked real parts of eigenfrequencies. While in regime (ii) (red shaded area), originally $\mathcal{APT}$S regime of the linear gyroscope is changed to $\mathcal{APT}$B regime in  nonlinear gyroscope, with unlocked real parts of eigenfrequencies. This nonlinearity-induced phase transition is reminiscent of similar effects in $\mathcal{PT}$ systems~\cite{lumer2013Nonlinearly,konotop2016Nonlinear}. In regime (iii) (yellow shaded area), both of the linear and nonlinear gyroscopes are $\mathcal{APT}$B and the real parts of their eigenfrequencies are unlocked. The corresponding frequency splittings are shown in Fig.~\ref{Fig3}(b), where the  frequency splitting of the nonlinear gyroscope is defined as $\Delta\omega_{\mathrm{NL}} =\mathrm{Re}[\omega^{\mathrm{NL}}_{+}-\omega^{\mathrm{NL}}_{-}]$.

Accordingly, nonlinear sensitivity enhancement versus $\Omega$ is plotted, see orange solid curve in Fig.~\ref{Fig3}(c). The nonlinear sensitivity enhancement is defined as  $\eta_{\mathrm{NL}} = \Lambda_{\mathrm{NL}}/\Lambda_{\mathrm{CONV}}$ with $\Lambda_{\mathrm{NL}} = \Delta\omega_{\mathrm{NL}}/\Delta\Omega$ and $\Lambda_{\mathrm{CONV}} = \Delta\omega_{\mathrm{CONV}}/\Delta\Omega$, where the frequency splittings $\Delta\omega_{\mathrm{NL}}$ and $\Delta\omega_{\mathrm{CONV}}$ are in respect to change of rotation speed $\Delta\Omega = \Omega - \Omega^{\mathrm{NL}}_{\mathrm{EP}}$. Linear sensitivity enhancement (blue dashed curve) is given twice for comparison.
It can be observed that $\mathcal{APT}$ gyroscope with and without nonlinearity both exhibit sensitivity enhancement in the vicinity of their own EPs. When experiencing a detectable change of rotation speed $0.01\,\mathrm{deg/s}$~\cite{lai2020Earth} from $\Omega^{\mathrm{NL}}_{\mathrm{EP}}$ in $\mathcal{APT}$B phase, the sensitivity enhancement of the nonlinear $\mathcal{APT}$ gyroscope can be up to $8.4\times10^3$ times that of the conventional gyroscope even with weak Kerr nonlinearity, while the linear gyroscope is locked at this point. At linear EP, although the frequency splitting of the nonlinear $\mathcal{APT}$ gyroscope is larger than that of linear device, it does not exhibit sensitivity enhancement since this point is far away from $\Omega^{\mathrm{NL}}_{\mathrm{EP}}$. When the rotation speed is even larger and far away from both linear and nonlinear EPs, there is no enhancement in these two gyroscopes.

\begin{figure}[tbp]
	\includegraphics[width=0.92 \columnwidth]{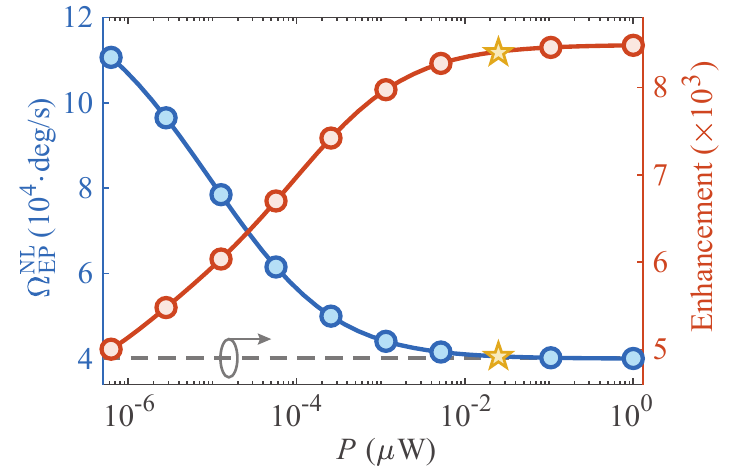}
	\caption{\label{Fig4} Rotation speed corresponding to nonlinear EP $\Omega^{\mathrm{NL}}_{\mathrm{EP}}$ and sensitivity enhancement in respect to a detectable change of rotation speed $0.01\,\mathrm{deg/s}$ from $\Omega^{\mathrm{NL}}_{\mathrm{EP}}$ versus the pump power $P$. The dashed curve indicates the sensitivity enhancement of linear $\mathcal{APT}$ gyroscope. The parameters used are the same as those in Fig.~\ref{Fig3}.}
\end{figure}

Finally, we show that tuning the location of EP of this nonlinear gyroscope can be readily realized by changing the pump power $P$, which can be used for ultrasensitive measurements of a wide range of rotation speed.  In Fig.~\ref{Fig4}, $\Omega^{\mathrm{NL}}_{\mathrm{EP}}$ is plotted versus pump power $P$ (blue curve), showing that the rotation speed at EP decreases when increasing pump power. The corresponding sensitivity enhancement $\eta_{\mathrm{NL}}$ (orange curve) is also shown, where the frequency splittings $\Delta\omega_{\mathrm{NL}}$ and $\Delta\omega_{\mathrm{CONV}}$ are in respect to a detectable change of rotation speed $0.01\,\mathrm{deg/s}$~\cite{lai2020Earth} from $\Omega^{\mathrm{NL}}_{\mathrm{EP}}$ in the $\mathcal{APT}$B regime. The sensitivity enhancement of the linear gyroscope is illustrated by dashed curve, which is independent on the pump power $P$.
Figure~\ref{Fig4} indicates that the sensitivity enhancement of the nonlinear gyroscope is always larger than that of the linear device, when tuning the pump power in a wide range. The case shown in Fig.~\ref{Fig3} is illustrated by yellow pentagrams in this figure. The presence of nonlinearity in this gyroscope shows great potential in practical applications when measuring a wide range of rotation speed.

\section{Conclusion}
In conclusion, we have proposed a nonlinear $\mathcal{APT}$ optical gyroscope and compared its performance with the linear device.
We find triple advantages for introducing Kerr nonlinearity into the $\mathcal{APT}$ system:
(i) the sensitivity can be enhanced with weak and experimentally accessible Kerr nonlinearity;
(ii) the measurable regime of weak rotations can be efficiently extended by breaking the $\mathcal{APT}$ symmetry due to the Kerr nonlinearity;
(iii) the EP position of the gyroscope can be continuously tuned in a wide range by changing the pump power, providing a feasible way to enhance the performance of the gyroscope on-situ.

Our work confirms that $\mathcal{APT}$ devices can serve as a powerful tool for highly sensitive rotation measurement and can be extended to study the role of nonlinearity in higher-order EP sensors.
Moreover, our proposal can be applied to other $\mathcal{APT}$ systems with various types of nonlinearity such as second-order nonlinearity~\cite{guo2016OnChip}, optomechanical interactions~\cite{aspelmeyer2014Cavity}, and hybridized atom-cavity interactions~\cite{thompson2013Coupling}.

\begin{acknowledgments}
Project supported by National Natural Science Foundation of China (Grants No.~11935006, No.~11774086, and No.~12064010), Science and Technology Innovation Program of Hunan Province (Grant No. 2020RC4047), Natural Science Foundation of Hunan Province of China (Grant No. 2021JJ20036), and Natural Science Foundation of Jiangxi Province of China (Grant No. 20192ACB21002).
\end{acknowledgments}

\textbf{Note added}---After submission of this paper, we notice a new experiment on nonlinear enhanced microresonator gyroscope~\cite{silver2021Nonlinear}, which demonstrated an optical gyroscope with a responsivity enhanced by a factor of around $10^{4}$ using the critical point of a spontaneous symmetry-breaking transition between counterpropagating light.


%

\end{document}